\documentclass{aa}  

\usepackage{graphicx}
\usepackage{txfonts}
\usepackage[dvipsnames]{xcolor}

\usepackage{soul}

\begin{document}

   \title{Density Fluctuations Associated with Turbulence and Waves}

   \subtitle{First Observations by Solar Orbiter}

\author{Yu. V. Khotyaintsev\inst{1}
          \and D. B. Graham\inst{1}
          \and A. Vaivads\inst{2}
          \and K. Steinvall \inst{1,3}
          \and N. J. T. Edberg\inst{1}
          \and A. I. Eriksson\inst{1}
          \and E.P.G. Johansson\inst{1}
          \and L. Sorriso-Valvo\inst{1}
          \and M. Maksimovic\inst{4}
          \and S.D. Bale\inst{5,6}
          \and T. Chust\inst{7}
          \and V. Krasnoselskikh\inst{8,5}
          \and M. Kretzschmar\inst{8,9}
          \and E. Lorf\`evre\inst{10}
          \and D. Plettemeier\inst{11}
          \and J. Sou\v{c}ek\inst{12}
          \and M. Steller\inst{13}
          \and \v{S}. \v{S}tver\'ak\inst{14}
          \and P. Tr\'avn\'i\v{c}ek\inst{5,12}
          \and A. Vecchio\inst{4,15}
          \and T. S. Horbury\inst{16}
          \and H. O'Brien\inst{16}
          \and V. Evans\inst{16}
          \and V. Angelini\inst{16}
          }
   \institute{Swedish Institute of Space Physics (IRF), Uppsala 75121, Sweden\\
              \email{yuri@irfu.se}
        \and
             Division of Space and Plasma Physics, School of Electrical Engineering and Computer Science,
KTH Royal Institute of Technology, Stockholm 11428, Sweden
\and
             Space and Plasma Physics, Department of Physics and Astronomy, Uppsala University, Uppsala 75120, Sweden
\and
LESIA, Observatoire de Paris, Universit\'e PSL, CNRS, Sorbonne Universit\'e, Univ. Paris Diderot, Sorbonne Paris Cit\'e, 5 place Jules Janssen, 92195 Meudon, France
\and
Space Sciences Laboratory, University of California, Berkeley, CA, USA
\and
Physics Department, University of California, Berkeley, CA, USA
\and
LPP, CNRS, Ecole Polytechnique, Sorbonne Universit\'e, Observatoire de Paris, Universit\'e Paris-Saclay, Palaiseau, Paris, France
\and
LPC2E, CNRS, 3A avenue de la Recherche Scientifique, Orl\'eans, France
\and
Universit\'e d'Orl\'eans, Orl\'eans, France
\and
CNES, 18 Avenue Edouard Belin, 31400 Toulouse, France
\and
Technische Universität Dresden, Helmholtz Str. 10, D-01187 Dresden, Germany
\and
Institute of Atmospheric Physics of the Czech Academy of Sciences, Prague, Czechia
\and
Space Research Institute, Austrian Academy of Sciences, Graz, Austria
\and
Astronomical Institute of the Czech Academy of Sciences, Prague, Czechia
\and
Radboud Radio Lab, Department of Astrophysics, Radboud University, Nijmegen, The Netherlands
\and
Imperial College London, South Kensington Campus, London SW7 2AZ, UK
\\
             }
   
   \date{Received September 15, 1996; accepted March 16, 1997}

 
  \abstract
   {}
   {The aim of this work is to demonstrate that the probe-to-spacecraft potential measured by RPW on Solar Orbiter can be used to derive the plasma (electron) density measurement, which both has a high temporal resolution and is of high accuracy. To investigate the physical nature of the solar wind turbulence and waves we analyze the density and magnetic field fluctuations around the proton cyclotron frequency observed by Solar Orbiter during the first perihelion encounter ($\sim$0.5~AU away from the Sun).}
   {We use the plasma density based on measurements of the probe-to-spacecraft potential in combination with magnetic field measurements by MAG to study fields and density fluctuations in the solar wind. In particular, we use the polarization of the wave magnetic field, the phase between the compressible magnetic field and density fluctuations, and the compressibility ratio (the ratio of the normalized density fluctuations to the normalized compressible fluctuations of B) to characterize the observed waves and turbulence.}
   {We find that the density fluctuations are $180^{\circ}$ out-of-phase (anti-correlated) with the compressible component of magnetic fluctuations for intervals of turbulence, while they are in phase for the circular-polarized waves. We analyze in detail two specific events with simultaneous presence of left- and right-handed waves at different frequencies. We compare observed wave properties to a prediction of the three-fluid (electrons, protons and alphas) model. We find a limit on the observed wavenumbers, $10^{-6} < k < 7 \times 10^{-6}$~m$^{-1}$, which corresponds to wavelength $7 \times 10^6 >\lambda > 10^6$~m. 
We conclude that most likely both the left- and right-handed waves correspond to the low-wavenumber part (close to the cut-off at $\Omega_{c\mathrm{He}++}$) proton-band electromagnetic ion cyclotron (left-handed wave in the plasma frame confined to the frequency range $\Omega_{c\mathrm{He}++} < \omega < \Omega_{c\mathrm{H}+}$) waves propagating in the outwards and inwards directions respectively. The fact that both wave polarizations are observed at the same time and the identified wave mode has a low group velocity suggests that the double-banded events occur in the source regions of the waves.}
   {}


   \maketitle
%

\section{Introduction}

The solar wind is abundant with plasma turbulence and waves \citep{belcher_large-amplitude_1971,tu_mhd_1995,bruno_solar_2013}. As collisions between particles are rare in the solar wind, the electromagnetic fluctuations play an important role in shaping the electron and ion velocity distribution functions \citep{marsch_solar_2018}. Identification of turbulence characteristics and of wave modes corresponding to the observed fluctuations is of primary importance for understanding the wave-particle interactions, and thus the electron and ion dynamics.

In addition to the extensively studied magnetic field and proton velocity \citep{bruno_solar_2013,sorrisovalvo_intermittency_1999}, density measurements also provide an important diagnostics for the identification and characterization of fluctuations. 
Turbulence in density fluctuations has been examined at fluid and kinetic scales, revealing interesting features such as power-law spectra \citep{chen_intermittency_2014}, intermittency \citep{hnat_compressibility_2005,carbone_arbitrary-order_2018,roberts_higher-order_2020}, multifractality \citep{sorriso-valvo_multifractal_2017}, and their radial evolution \citep{bruno_radial_2014}. The characteristics of density turbulence strongly depend on the nature of the solar wind. For nearly incompressible (typically fast) Alfv\'enic solar wind, density fluctuations are mostly passively advected by magnetic and velocity fields, which dominate the dynamics \citep{goldreich_toward_1995,chen_density_2012}. In more compressible (typically slow) solar wind, density fluctuations are not simply passively advected by magnetic and velocity fluctuations, but are rather actively contributing to the nonlinear cascade \citep{hadid_energy_2017}, with enhanced turbulent signatures \citep{bruno_radial_2014}. 
It is often found that the compressible fluctuations of the magnetic field are anti-correlated with density, and thus such fluctuations are interpreted as pressure-balanced structures \citep{yao_multi-scale_2011}. Such compressible fluctuations have been attributed to the kinetic \citep{howes_slow-mode_2012} and MHD \citep{verscharen_kinetic_2017} slow mode. 

Low-frequency waves are also commonly observed in the solar wind. One type of wave commonly found during intervals of the predominantly radial magnetic field are circularly polarized electromagnetic waves at frequencies close to the proton-cyclotron frequency, $f_{cp}$ \citep{jian_observations_2010,bale_highly_2019}. Such waves have very small wave-normal angles with respect to the background field and are often observed in extended bursts lasting from several to several tens of minutes \citep{jian_electromagnetic_2014,boardsen_messenger_2015}.
One of the possible energy sources for the growth of such waves is the ion temperature anisotropy \citep{davidson_electromagnetic_1975}, which is supported by the correlation between the transverse wave power close to $f_{cp}$ and the proton perpendicular temperature anisotropy \citep{bourouaine_correlations_2010}.  
These waves are often interpreted as electromagnetic ion cyclotron waves, which have intrinsic left-handed polarization in the plasma frame. However, based solely on the magnetic field measurement it is not possible to determine the wave polarization in the plasma frame. This presents a challenge since the wave polarization observed in the spacecraft frame may be modified due to the Doppler shift in a fast-flowing solar wind. \citet{bowen_electromagnetic_2020} have used electric field measurements to identify the sense of wave propagation and from this, the plasma frame polarization. They found waves with both senses of the plasma frame polarization. However, the approach using the electric field requires accurate calibration of the electric field gain ($\delta E/E < V_A/V_{\mathrm{sw}}$), which is very challenging to achieve given the relatively short electric field antennas on both Parker Solar Probe and Solar Orbiter. Therefore, we take a different approach and will use density fluctuations to identify the wave mode corresponding to the observed waves.

When designing electric field measurements for the Solar Orbiter mission \citep{muller_solar_2020} it was important to include the capability to measure high-quality electric fields and density fluctuations up to frequencies of at least about 100 Hz \citep{vaivads_low-frequency_2007}. This is of high importance for studying plasma processes and identifying plasma waves in the ion and electron kinetic range. Particularly, including the capability to current bias antennas reduces the noise level and increases the accuracy of electric field measurements at those frequencies, and in addition, it allows the use of satellite potential estimates as a proxy for fast plasma density measurements.  

The main purpose of this paper is to demonstrate that the probe-to-spacecraft potential measured by RPW \citep{maksimovic_solar_2020} on Solar Orbiter can be used to derive the plasma density measurement, which has both a high temporal resolution and is of high accuracy. We also demonstrate that such measurements provide a valuable diagnostic of plasma waves and turbulence. First, we present the procedure for deriving the density from RPW measurements. Then we use the obtained density together with magnetic field measurements by MAG \citep{horbury_solar_2020} to study density fluctuations associated with turbulence and waves in the solar wind.

\section{Calibration of plasma density}

   \begin{figure}
   \centering
   \includegraphics[width=\hsize]{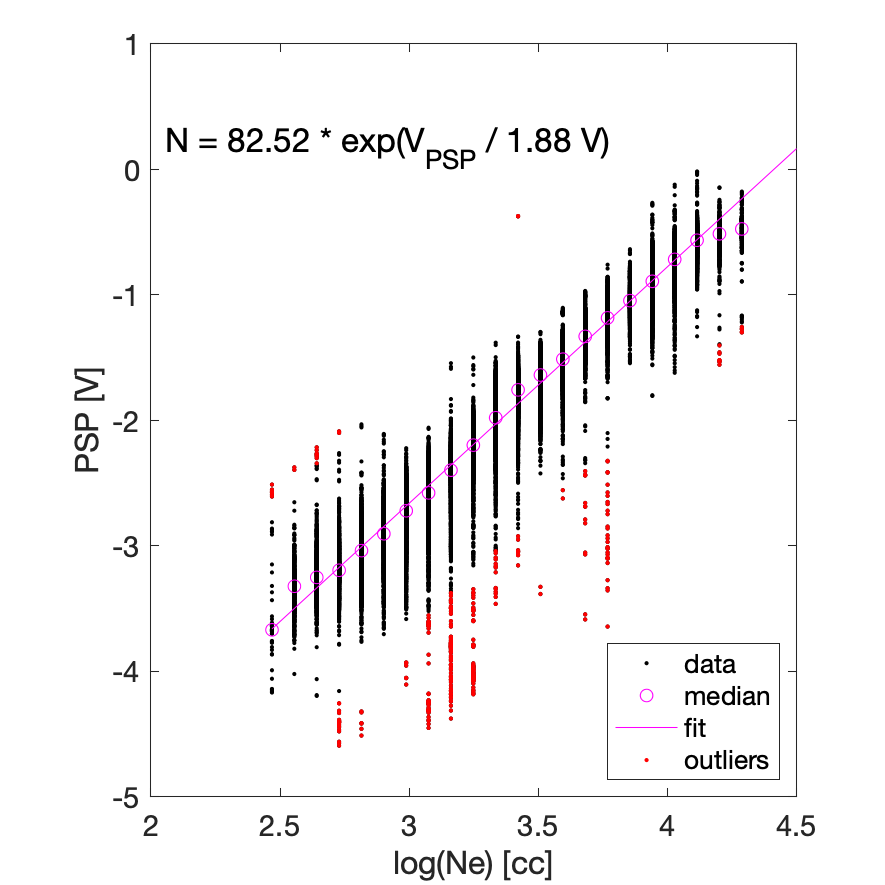}
      \caption{Fit of $V_{\mathrm{PSP}}$ to $\log(n_e)$, where $n_e$ is the QTN-based density. The black dots show data points used for the linear fit (pink line), and red dots show points that were marked as outliers. The pin circles show the median values for each of the $n_e$ values. The data used in these plots corresponds to the time interval from May 30 to August 11 2020, with most of the data points coming from June 2020. 
              }
         \label{fig:Ne_Vpsp}
   \end{figure}

Radio and Plasma Waves (RPW) instrument \citep{maksimovic_solar_2020} on Solar orbiter has three electrical antennas. A current bias is applied to each of the antennas Each of the antennas brings the antennas closer to the local plasma potential. This enables sensitive measurement of the DC \citep{steinvall_solar_2021} and low-frequency electric fields \citep{chust_observations_2021, kretzschmar_whistler_2021} and the spacecraft potential. 

First, we establish a relation between the probe-to-spacecraft potential, $V_{\mathrm{psp}}$ and the electron density as commonly done in space plasmas \citep[e.g.][]{pedersen_electron_2008}. 
The spacecraft floating potential is reached when the total current to the spacecraft is zero, i.e. the photo-electron emission from the spacecraft, $I_{\mathrm{ph}}$, is balanced by the plasma electron current, $I_{\mathrm{e}}$, to the spacecraft:
 \begin{equation}
    I_{\mathrm{e}} + I_{\mathrm{ph}} = 0\,.
    \label{CurrentBalance} 
 \end{equation}
Here we consider only the major current contributors and neglect the smaller contribution from e.g. plasma ions and secondary emission. Using the following notation:
   \[
      \begin{array}{lp{0.8\linewidth}}
         I_{e}  & plasma electron current     \\
         I_{\mathrm{ph}}               & photo-electron current                    \\
         r_0             & unperturbed zone radius             \\
         \rho_0          & unperturbed density in the zone     \\
         T_e             & electron temperature \\
         T_{\mathrm{ph}}          & photo-electron temperature              \\
         V_{\mathrm{1..3}}          & probe-to-spacecraft potential the tree probes              \\
         V_{\mathrm{PPL}}          & local probe-to-plasma potential            \\
         V_{\mathrm{PSP}}          & probe-to-spacecraft potential averaged from the three probes              \\
         V_{\mathrm{SC}} & potential of the spacecraft with respect to plasma at a large distance
      \end{array}
   \]
   \begin{figure*}
   \centering
   \includegraphics[width=\hsize]{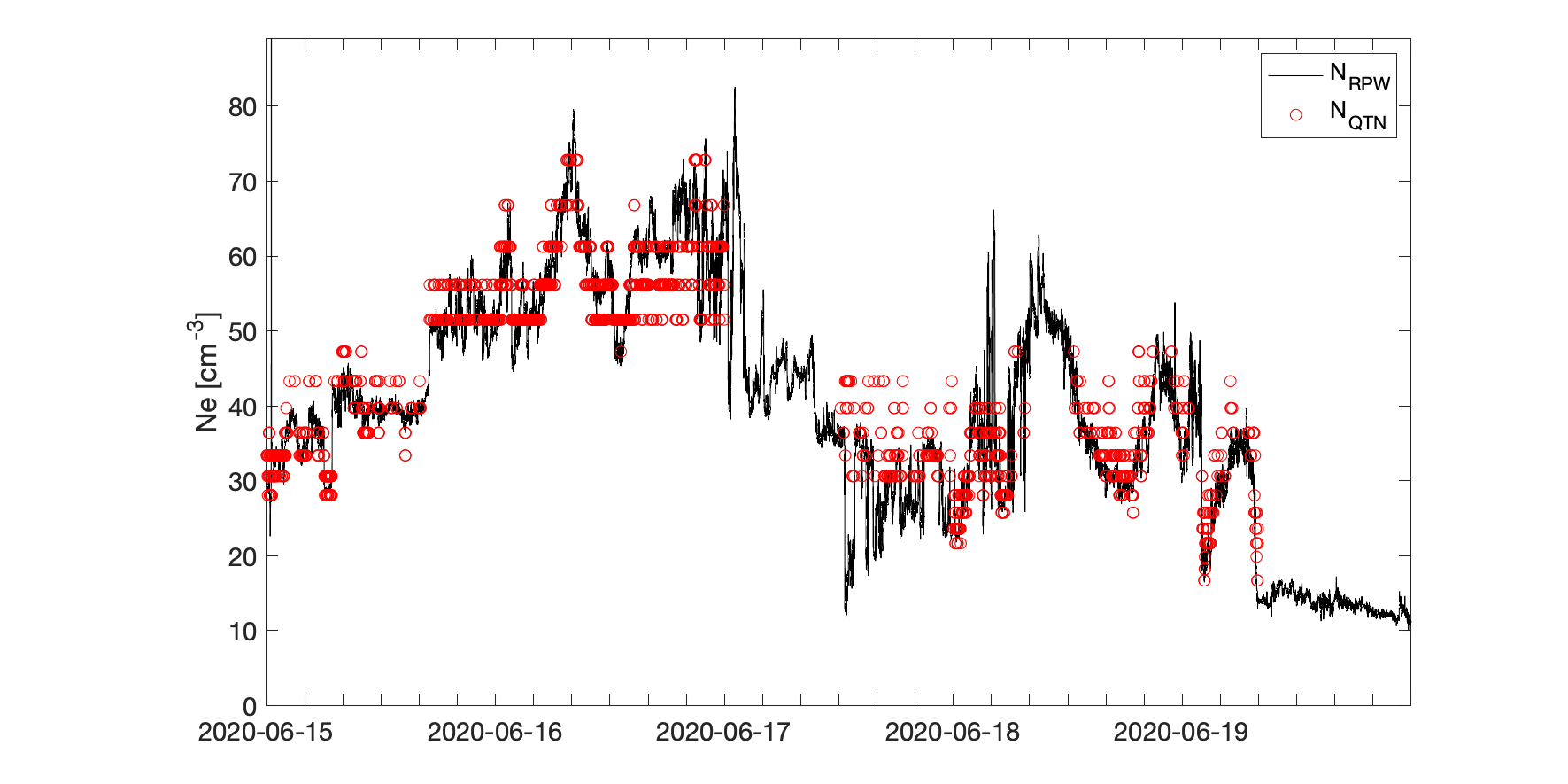}
   \caption{Comparison of $N_{\mathrm{RPW}}$ (BIAS) to the QTN-based density (red circles) for a 6-day interval starting on July 15 2020.
              \label{fig:Ne_Vpsp_Example}}%
    \end{figure*}
   the currents in Eq.~\ref{CurrentBalance} for a single photo-electron population are given by \citep{pedersen_solar_1995}:
\begin{eqnarray}
  I_{\mathrm{e}} & = & -e n_e S \left( \frac{k_B T_e}{2 \pi m_e} \right)^{1/2} \left( 1 + \frac{e V_{\mathrm{SC}}}{k_B T_e} \right) \\
  I_{\mathrm{ph}} & = & I_{\mathrm{ph0}} \exp \left( - \frac{e V_{\mathrm{SC}}}{k_B T_\mathrm{ph}} \right) \,,
   \end{eqnarray}
where $S$ is the total surface area of the spacecraft body. As the RPW probes are located relatively close to the spacecraft, the local plasma potential at the probe location will have a significant contribution from the spacecraft potential. Therefore, the potential difference between the spacecraft and the local plasma potential at the location of the probes will correspond only to a fraction of the spacecraft potential
\begin{equation}
   V_{\mathrm{PSP}} + V_{\mathrm{PPL}} = - \alpha V_{\mathrm{SC}} \,,
\end{equation}
where $\alpha < 1$, and we have also included a local probe-to-plasma potential, $V_{\mathrm{PPL}}$, which is of the order of 1 V for a biased probe and is approximately constant. Using the current expressions above and assuming $e V_{\mathrm{SC}} \ll k_B T_e$ we can find an approximate dependence of $n_e$ on $V_{\mathrm{PSP}}$
\begin{equation}
   n_e \simeq N_{\mathrm{RPW}} = N_0 \exp \left( \frac{ V_{\mathrm{PSP}}}{\beta} \right) \,,
   \label{eq:ne_Vpsp}
\end{equation}
where $\beta$ is proportional to $T_{\mathrm{ph}}$. We note that in particular for the magnetospheric missions it is usually necessary to use two photo-electron populations with different temperatures, \citep{pedersen_electron_2008,graham_enhanced_2018}, as the spacecraft are crossing a wide range of plasma environments and the variation of the spacecraft potential is large, i.e. from several volts in the solar wind and up to $\sim$100 V in  the magnetospheric lobes. In the case of Solar Orbiter it is sufficient with a single population as the spacecraft stays in a relatively stable environment in comparison to magnetospheric missions. 

We obtain $V_{\mathrm{PSP}}$ by combining the individual probe voltages $V1$, $V2$ and $V3$ which are measured by RPW. We cannot simply take an average of the three probes as they are located differently with respect to the spacecraft and have slightly different photo-emission characteristics. We first remove the offset between $V2$ and $V3$, compute an average between these two probes, and then scale-up the average to match $V1$, i.e. accounting for the fact that $V1$ is located further away from the electrostatic center of the spacecraft than $V2$ and $V3$~\citep{steinvall_solar_2021}. Finally, we average this scaled-up quantity with $V1$ to give $V_{\mathrm{PSP}}$.

The coefficients $N_0$ and $\beta$ in Eq. \ref{eq:ne_Vpsp} can be determined empirically by fitting $V_{\mathrm{PSP}}$ to the reference plasma density data. RPW provides a sensitive measurement of the plasma quasi-thermal noise (QTN). When the QTN signal is of sufficient strength it is possible to identify the spectral peak at the electron plasma frequency, and then derive the plasma density. We will use this density as the reference for our fitting. Fig. \ref{fig:Ne_Vpsp} shows an example of such a fit for the time interval from May 30 to August 11 2020. The black and red dots show the data, with the red dots being marked as the outliers (0.8\% of the dataset) and excluded from the fitting. The discrete distribution of the data points in $\log(n_e)$ is related to the frequency resolution of the receiver providing the QTN measurements. The pink circles show the median value for a particular density. Finally, the pink line shows the least-squares linear fit. For this dataset we obtain $N_0 = 81.5$ $\mathrm{cm}^{-3}$ and $\beta=1.88$ V, and one can see that the fit is of good quality as the median values lie very close to the line.

 \begin{figure*}
   \centering
   \includegraphics[width=\hsize]{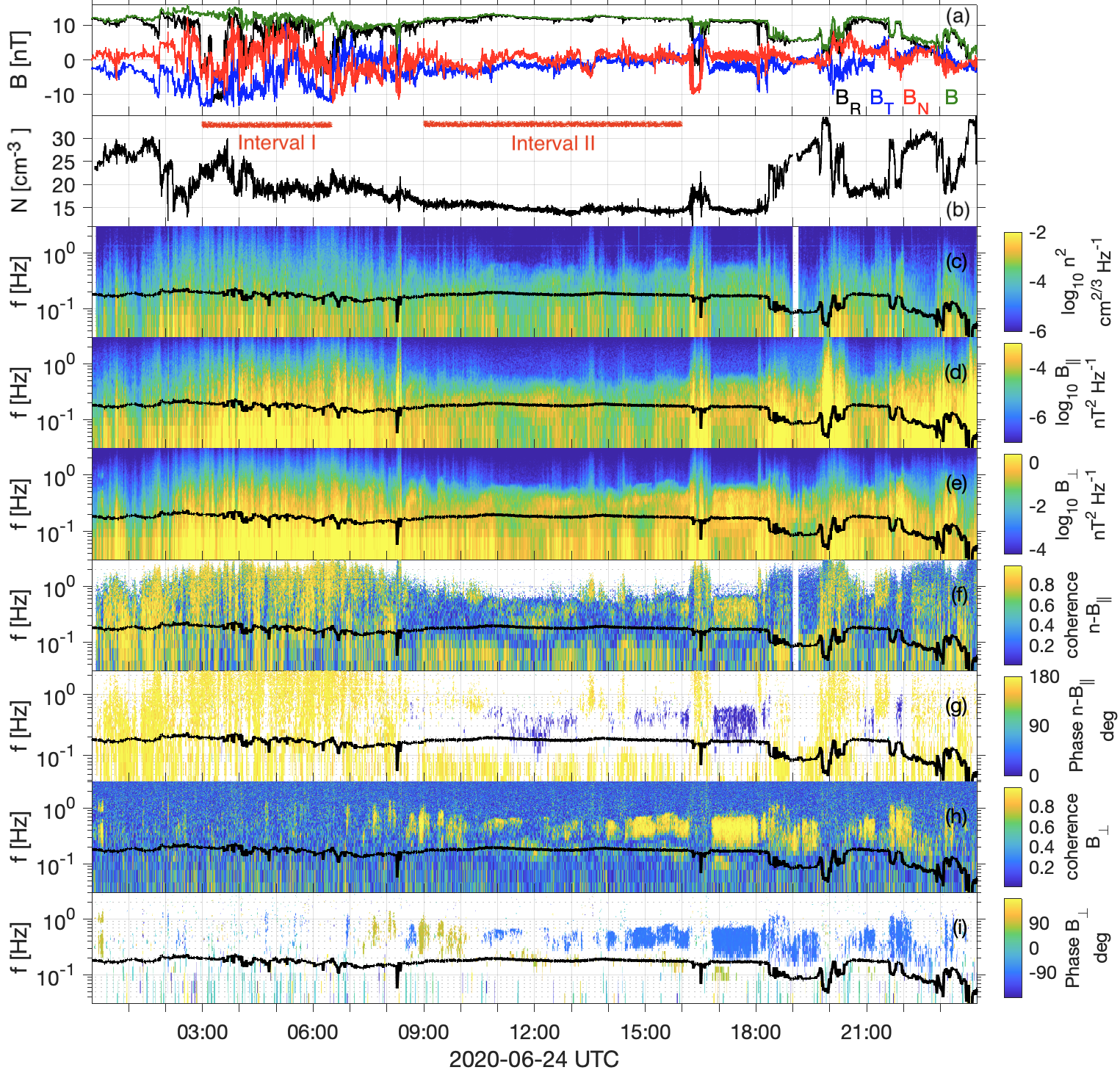}
   \caption{Event overview. The event contains 2 specific intervals which we analyze in detail: interval I is dominated by turbulence and interval II with abundant coherent waves. Panels from top to bottom show: (a) magnetic field vector in RTN coordinates, (b) density $n_e$, (c-e) power spectrum of density, $B_{||}$ and $B_{\perp}$ fluctuations, (f) coherence and (g) phase between $n_e$ and $B_{||}$, (h) coherence and (i) phase between the two $B_{\perp}$ components The black lines in panels (c)-(i) show the proton cyclotron frequency, $f_{cp}$.
              \label{fig:Event_overview}}%
    \end{figure*}

As the $V_{\mathrm{PSP}}$ data is available nearly continuously (sampling frequency of 16 or 256 samples/sec) we can use Eq. \ref{eq:ne_Vpsp} with the coefficients obtained from the fitting to get a continuous plasma density $N_{\mathrm{RPW}}$. To verify the results we plot the density obtained by this method together with the QTN-based density in Fig. \ref{fig:Ne_Vpsp_Example} for a 5-day interval starting on June 15, 2020. We can see that the QTN-based density has a limited resolution and that sometimes the derived plasma peak jumps between the nearby frequency bins, and sometimes artificial interference lines can be confused with a natural plasma line. But we can relatively easily identify the problematic intervals, and find a generally excellent agreement between the two datasets even at the smaller temporal scales. From this we conclude that $N_{\mathrm{RPW}}$ provides an accurate density measurement. 

We perform the fitting procedure described above on the time intervals of several weeks to two months. The need to split into shorter intervals is related to the major bias current changes, as such changes introduce step-like changes into $V_{\mathrm{PSP}}$. To avoid having an artificial discontinuity in the resulting density, the data on the two sides of the discontinuity needs to be fitted separately. The bias current needs to be changed by a telecommand in order to follow the photoemission of the probes, which is needed for optimal electric field measurements. The photoemission depends on the distance between Solar Orbiter and the Sun, which is changing significantly along the orbit. Thus, the bias current needs to be changed with intervals of several weeks to two months. By applying the fitting procedure to the time intervals which can be well-fitted by a single exponent, we obtain a set of calibration coefficients $N_0$ and $\beta$ which is then used to produce $N_{\mathrm{RPW}}$, which is made publicly available as an L3 data product. 

\section{Density fluctuations}
   \begin{figure*}
   \centering
   \centering\includegraphics[scale=0.6]{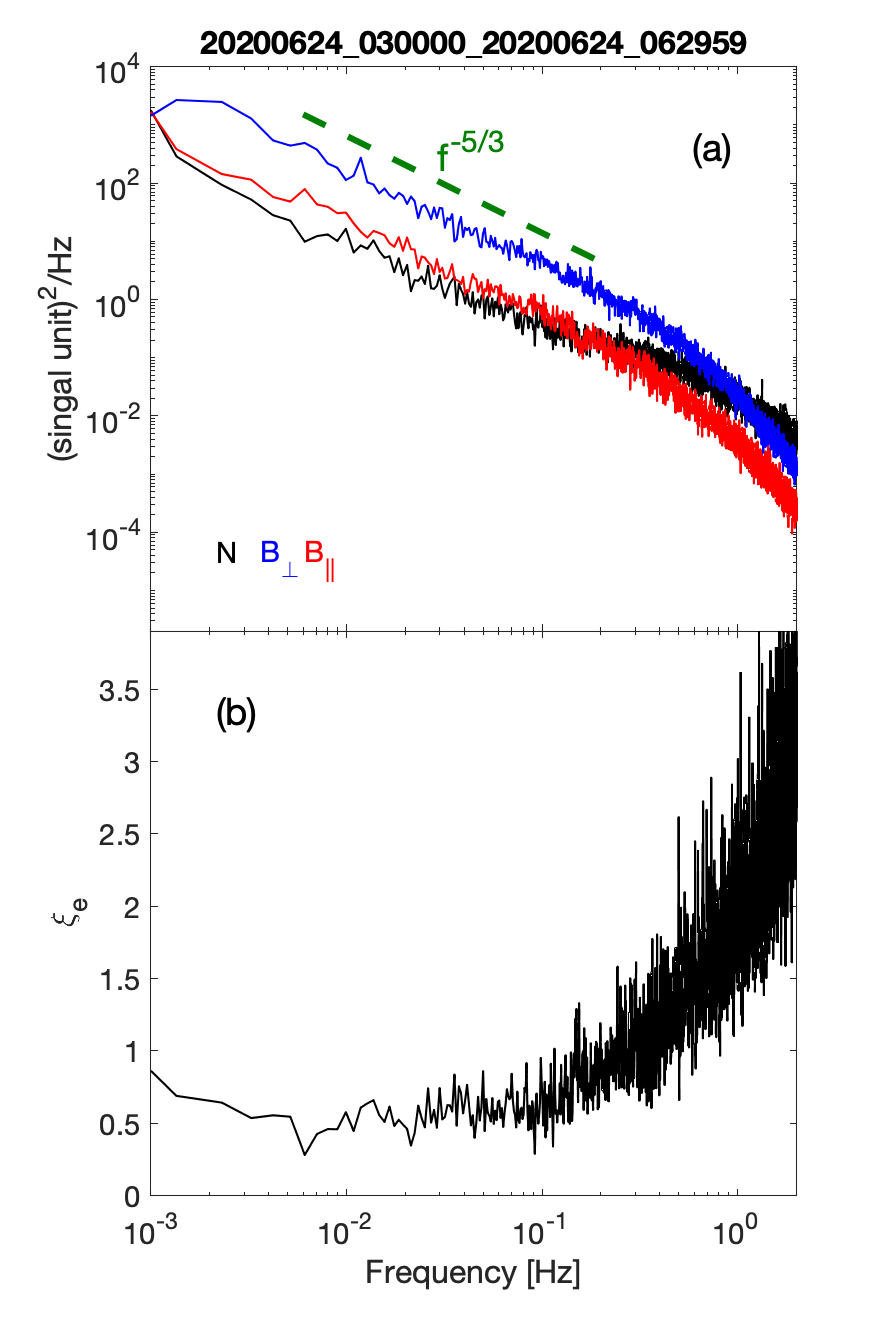}
   \centering\includegraphics[scale=0.6]{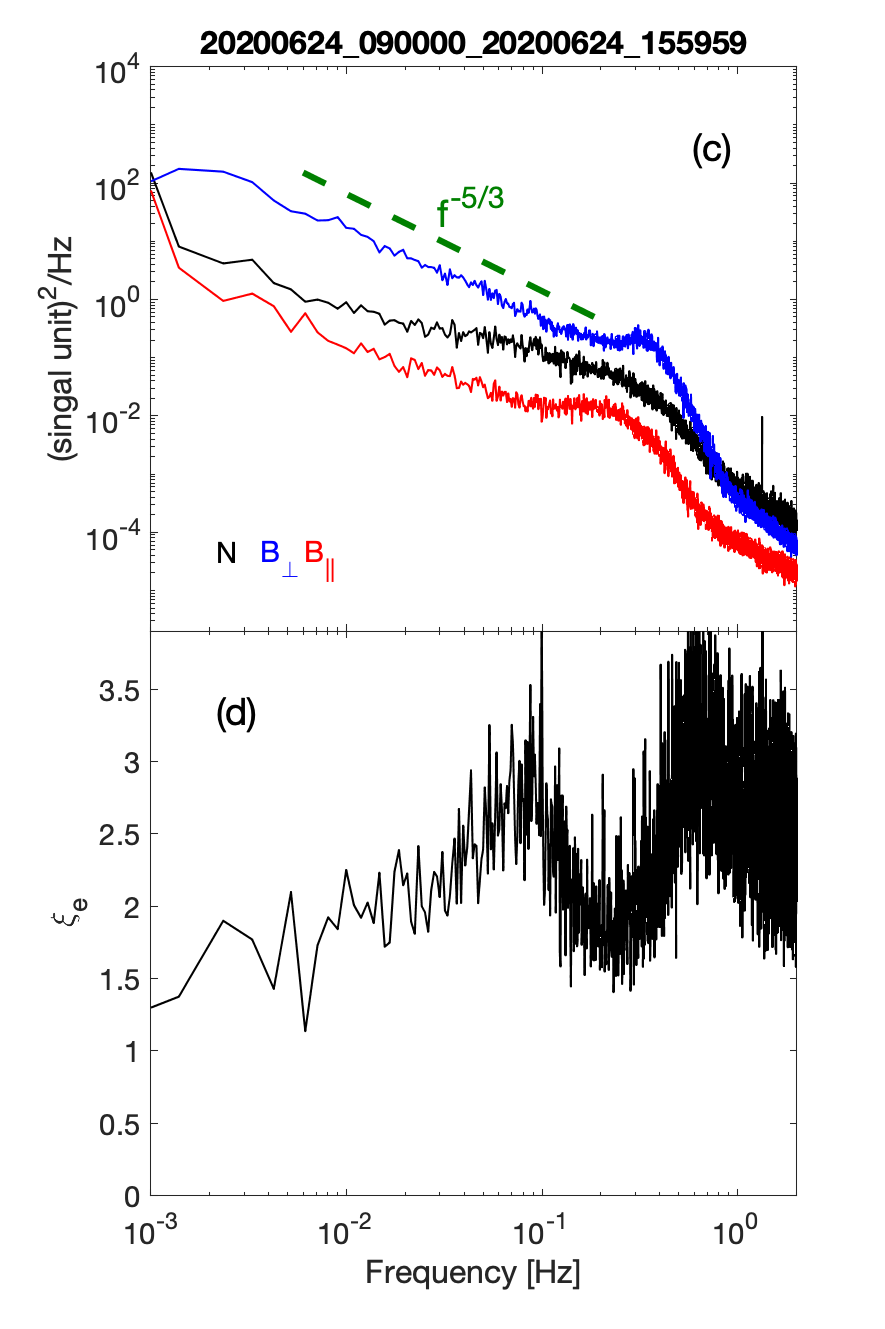}
   \caption{(a,c) Spectra of $n_e$, $B_{\perp}$ and $B_{||}$ and (b,d) compressibility ratio $\xi_e = (\delta n/n)/(\delta B_{||}/B)$ for intervals I and II (see Figure\ref{fig:Event_overview}b). The green dashed lines in (a,c) show the slope of $\sim f^{-5/3}$.
              \label{fig:Power_Spectra}}%
    \end{figure*}

$V_{\mathrm{PSP}}$ is available with a sampling frequency of 16 or 256 samples/sec. As the characteristic time for charging of the spacecraft following a change in the plasma environment (given by $\tau = RC$, where $R$ and $C$ are spacecraft sheath resistance and capacitance) is shorter than 1 ms, we can use $N_{\mathrm{RPW}}$ to study plasma density fluctuation up to the corresponding Nyquist frequencies. To verify that $N_{\mathrm{RPW}}$ exhibits a correct physical behavior we compare the observed fluctuations in the magnetic field. We focus on the data from the first perihelion encounter by Solar Orbiter in June 2020 when the spacecraft was located at $\sim$0.5~AU from the Sun. At this time only the 16 samples/sec data was available for $V_{\mathrm{PSP}}$.

To illustrate some typical types of magnetic field and density fluctuations observed in the solar wind we use the event on June 24, 2020. The overview of the event is shown in Fig. \ref{fig:Event_overview}. Panel (a) shows B in RTN coordinates. This event contains both (I) an interval of a highly varying magnetic field between 01:40 and 08:30 UT, and (II) a long interval of a rather constant radial magnetic field between 08:30 and 18:00 UT. Interval (II) has a rather constant density $n\sim$15 cm$^{-3}$, while there are significant density variations during interval (I). Panels (c)-(e) show the power spectrograms of $n$, $B_{||}$ and $B_{\perp}$, where the parallel and perpendicular components are defined with respect to the background magnetic field (B low-pass filtered at 0.01 Hz). We can see that interval (I) has a broadband turbulent spectrum of fluctuations, while during interval (II) we can identify a more narrow-band spectrum with the peak above the proton cyclotron frequency. Panel (f) shows the coherence between the fluctuations of $n$ and $B_{||}$. The coherence between two signals is close to 1 for highly correlated signals and is close to 0 when the signals are not correlated to each other \citep{means_use_1972}. When the coherence is high (above 0.7) we can also compute the phase $\phi$ between $\delta n$ and $\delta B_{||}$  shown in panel (g). We can see that $\delta n$ and $\delta B_{||}$ are in anti-phase ($\phi = 180^{\circ}$) most of the time. Such behavior is expected for pressure-balanced structures. For the narrow-band waves, on the other hand, $\delta n$ and $\delta B_{||}$ are in phase ($\phi = 0^{\circ}$), so these are fluctuations likely exhibit changes in the total pressure if we assume that the temperature is approximately constant. We note that both types of behavior (in- and out-of-phase) make physical sense as they correspond to the behavior expected for pressure-balanced structures vs compressible waves, which suggests that the density fluctuations are measured correctly.

In panel (h) of Fig. \ref{fig:Event_overview} we show the cross-coherence between the two perpendicular components of B, which is high for the narrow-band waves and close to zero elsewhere. In panel (i) we also show the phase between the two components which provides the sense of polarization of the waves. We can see that the waves are mostly left-hand polarized ($-90^{\circ}$), but some patches of right-hand polarization also exist. 
We also note that between 10:30 and 17:30 UT there is a different sense of polarization at different frequencies, i.e. left-handed at high frequencies and right-handed at low frequencies (close to or below $f_{cp}$ in this case). 
We have also applied the Singular Value Decomposition (SVD) technique \citep{santolik_singular_2003,taubenschuss_wave_2014} to further characterize polarization of the right- and left-handed waves (not shown), and we find that these waves have polarization close to circular, and very small wave normal angles ($\theta_{kB} < 10^{\circ}$). 
Later in this paper we will look at these waves in detail and attempt to identify the wave-mode based on the observed wave characteristic, in particular on the relative power in the $\delta n$ and $\delta B_{||}$.

   \begin{figure*}
   \centering
   \includegraphics[width=18cm]{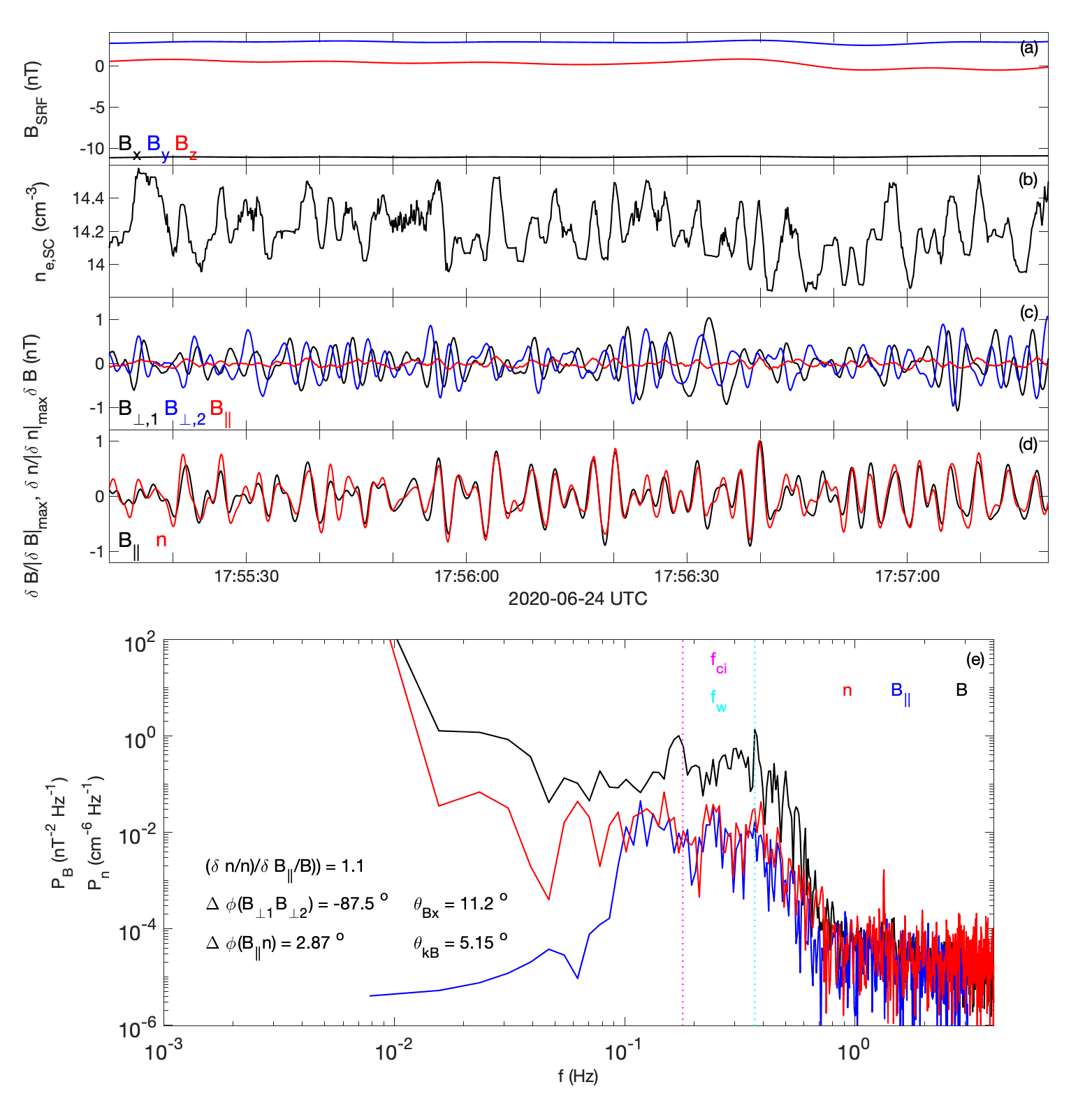}
   \caption{Examples of a quasi-circularly-polarized wave. (a) the background magnetic field in SRF coordinates, (b) density $n_e$, (c) wave magnetic field $\delta B$ in field-aligned coordinates, (d) normalized compressible magnetic component $\delta B_{||}/B$ and density $\delta n/n$, (e) power spectra of $n_e$, $B_{\perp}$ and $B_{||}$. }
         \label{fig:Example_wf_waves}%
    \end{figure*}

In Fig. \ref{fig:Power_Spectra} we compare the power spectral density for the two intervals discussed above, which are (I) dominated by turbulence and (II) contains a combination of turbulence and quasi-circularly-polarized waves. 
Panel (a) shows the power spectra for fluctuations of $n$, $B_{\perp}$ and $B_{||}$. The three spectra show a clear power-law dependence over a broad frequency range, and spectral exponents compatible with typical turbulence Kolmogorov scaling $\sim f^{-5/3}$ \citep[see e.g.][]{bruno_solar_2013}. 
The power in transverse fluctuations exceeds the compressible power by an order of magnitude, which is typical for the solar wind \citep{belcher_large-amplitude_1971,howes_slow-mode_2012,kiyani_enhanced_2012}. 
The spectrum of density fluctuations nicely follows the parallel magnetic spectrum up to $\sim0.1$ Hz, where the density spectrum becomes flatter. In panel (b) we plot the ratio $\xi_e$ between the normalized amplitude of density fluctuations to the normalized amplitude of compressible magnetic fluctuations \citep{verscharen_kinetic_2017}:
\begin{equation}
   \frac{\delta N}{N_0} = \xi_e \frac{\delta B_{||}}{B_0} \,.
\end{equation}
We can see that the compressibility ratio $\xi_e$ is close to 0.5 for frequencies below 0.1 Hz, and then increases for higher frequencies. These values are within the expected range for solar wind turbulence \citep{howes_slow-mode_2012}. Overall, we observe a good agreement between the magnetic and density spectra, i.e. the spectra follow each other and there are no unexpected features/structures. This indicates that the density spectrum based on $N_{\mathrm{RPW}}$ provides a good measurement of the density turbulence.

Figure \ref{fig:Power_Spectra}c shows spectra for the time interval (II), with a high abundance of waves. First, we note that for the perpendicular magnetic field component, the power-law dependence for this interval is the same as for interval (I), with the spectral exponent close to the Kolmogorov value. However, the power is an order of magnitude lower than for interval (I). 
We also note that parallel fluctuations are much weaker than for interval (I), i.e. their power is two orders of magnitude below the transverse power. Additionally, they have a slightly shallower spectrum. 
The spectrum of density fluctuations is now flatter, with scaling close to $f^{-1}$. 
This can be clearly seen in panel (d) which shows the compressibility ratio $\xi_e$. The ratio is increasing with frequency and reaches 2.5 at 0.1 Hz. Above 0.1 Hz we see an end to the power-law behavior, which is caused by a spectral peak in both transverse and compressible magnetic components. This peak corresponds to the quasi-circularly polarized waves. 

Despite the fact that the two intervals we considered have similar power-law dependence for the transverse magnetic power (albeit the power is lower for interval II), they clearly show sensibly different behavior of the density spectrum. 
Detailed analysis of density fluctuations additionally shows that interval (I) also features standard intermittent density fluctuations that are characteristic of turbulence, while for interval (II) the density fluctuations show no intermittency \citep[not shown, see][for details on intermittency in these samples]{carbone_statistical_2021}. This observation, together with the $f^{-1}$ spectral dependence, may indicate that interval (II) contains more fast and Alfv\'enic wind, where the turbulence has not fully developed yet \citep{bruno_radial_2003}. This interval instead shows a high abundance of circularly-polarized waves close to the proton-cyclotron frequency, likely generated by kinetic ion instabilities \citep{marsch_solar_2018}. 
In this interval, and in similar regions, such waves likely have the dominant impact on ion dynamics and heating, while the turbulence is still not fully developed. 
These types of fluctuations, associated with faster wind, can be similar to the solar wind source regions closer to the Sun. 
It is therefore extremely interesting to study the nature of the circularly-polarized waves in more detail.

\section{Quasi-circular waves}

Fig. \ref{fig:Example_wf_waves} shows an example of a time series of such waves. Panel (a) shows the background \textbf{B} for reference and panel (b) shows the density. Panel (c) shows the B-fluctuations in field-aligned coordinates and one can see that $\delta B_{\perp} \gg \delta B_{||}$, and that $\delta B_{\perp 1}$ is shifted with respect to $\delta B_{\perp 2}$ by approximately a quarter-wave-period which corresponds to left-handed polarization. We also find $\theta_{kB} = 5^{\circ}$ for this event. Panel (d) shows the normalized $\delta n$ and $\delta B_{||}$ and one can clearly see that they are in phase, and otherwise almost identical to each other. We find on average $\xi_e = (\delta n/n)/(\delta B_{||}/B) = 1.1$. Panel (e) shows the spectra corresponding to the time-series, and we note even the excellent agreement between the spectra of $\delta n$ and $\delta B_{||}$ in the frequency range between 0.1 and 1 Hz. At higher frequencies the spectrum is likely reaching the noise floor.


We search for solar wind intervals of coherent low-frequency quasi-circularly polarized waves using the following procedure: We divide the continuous magnetic field data into segments of 1024 points. We divide ${\bf B}$ into the background and fluctuating components $\delta {\bf B}$ by low-pass filtering below $0.1$~Hz and band-pass filtering between $0.1$~Hz and $3$~Hz, respectively. We rotate $\delta {\bf B}$ into field-aligned coordinates and perform minimum variance analysis to determine the minimum 
variance direction (equivalently the wave vector ${\bf k}$ direction) and 
the eigenvalues $\lambda_{1,2,3}$ of the maximum, intermediate, and minimum variance directions. 

Using data from June 2020 we identify coherent wave intervals as segments satisfying the following criteria: 
(1) The peak amplitude of $\delta {B}$ exceeds $0.1$~nT. 
(2) The angle between the minimum variance direction and ${\bf B}$, $\theta_{kB}$, is less than $25^{\circ}$. 
(3) $\lambda_1/\lambda_2 < 2$.
(4) $\lambda_2/\lambda_3 > 5$. 
(5) The power spectrum has a spectral peak between $0.1$~Hz and $3$~Hz. 
(6) The average phase difference between the two perpendicular components of $\delta {\bf B}$ is between $70^{\circ}$ and $110^{\circ}$. 
(7) The phase difference between $\delta n$ and $\delta B_{\parallel}$ is less than $45^{\circ}$. 

\begin{figure}
   \centering
   \includegraphics[width=\hsize]{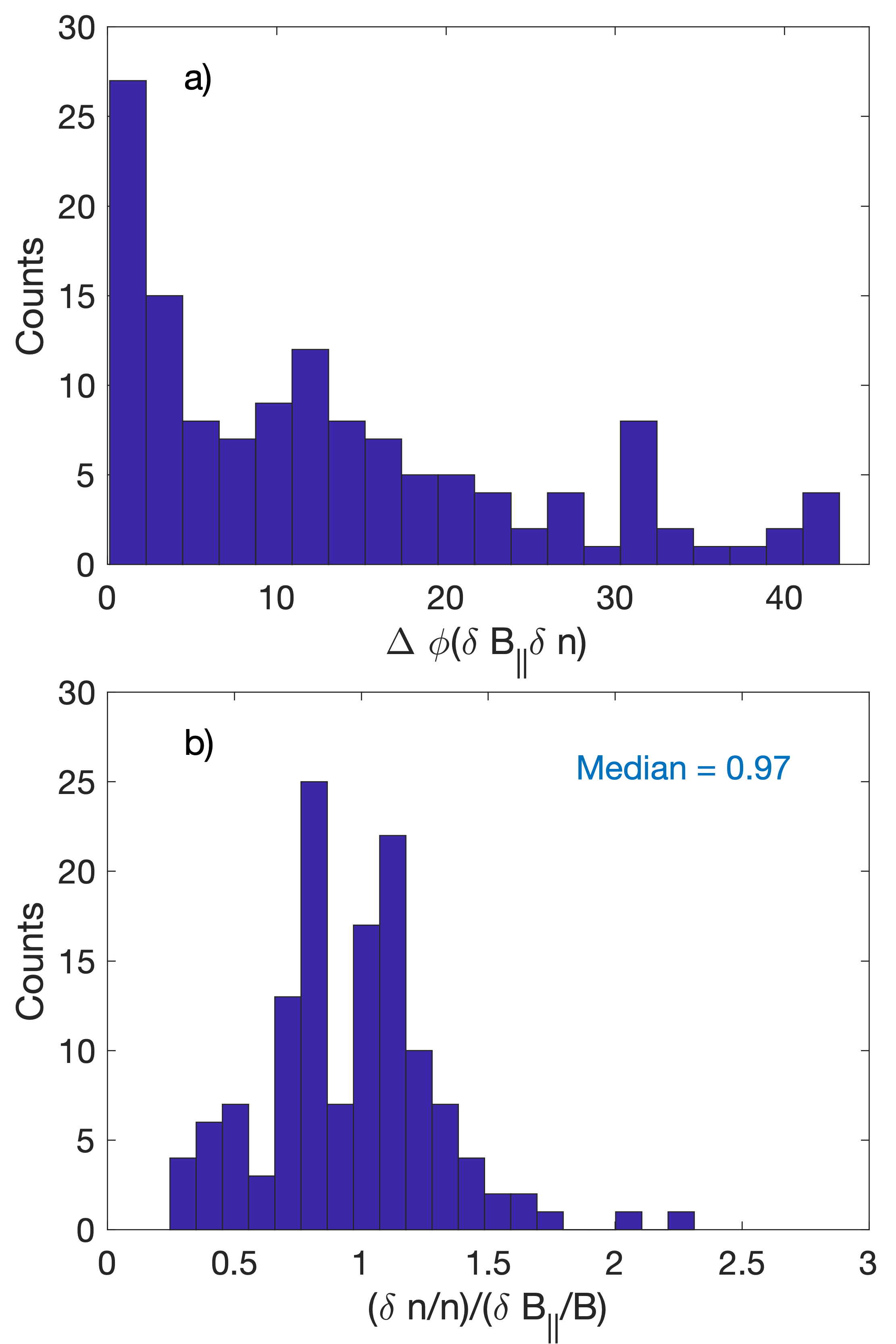}
      \caption{Statistics for quasi-circularly-polarized waves (see text for the selection criteria). (a) phase difference $\delta n$ and $\delta B_{\parallel}$ (b) compressibility ratio $\xi_e$. 
              }
         \label{fig:Wave_stats}
   \end{figure}

We find 132 segments satisfying these criteria. Figure \ref{fig:Example_wf_waves} shows an example of one of these segments and Figure \ref{fig:Wave_stats} shows the statistical results. We find that $\delta n$ and $\delta B_{||}$ are close to being in phase, with only a minor fraction of the events for which the phase difference between the two quantities exceeds $20^{\circ}$ (panel a). Also we find a rather narrow distribution of $\xi_e = (\delta n/n)/(\delta B_{||}/B)$ centered around 1. 

\section{Wave theory}
   \begin{figure*}
   \centering
   \includegraphics[width=\hsize]{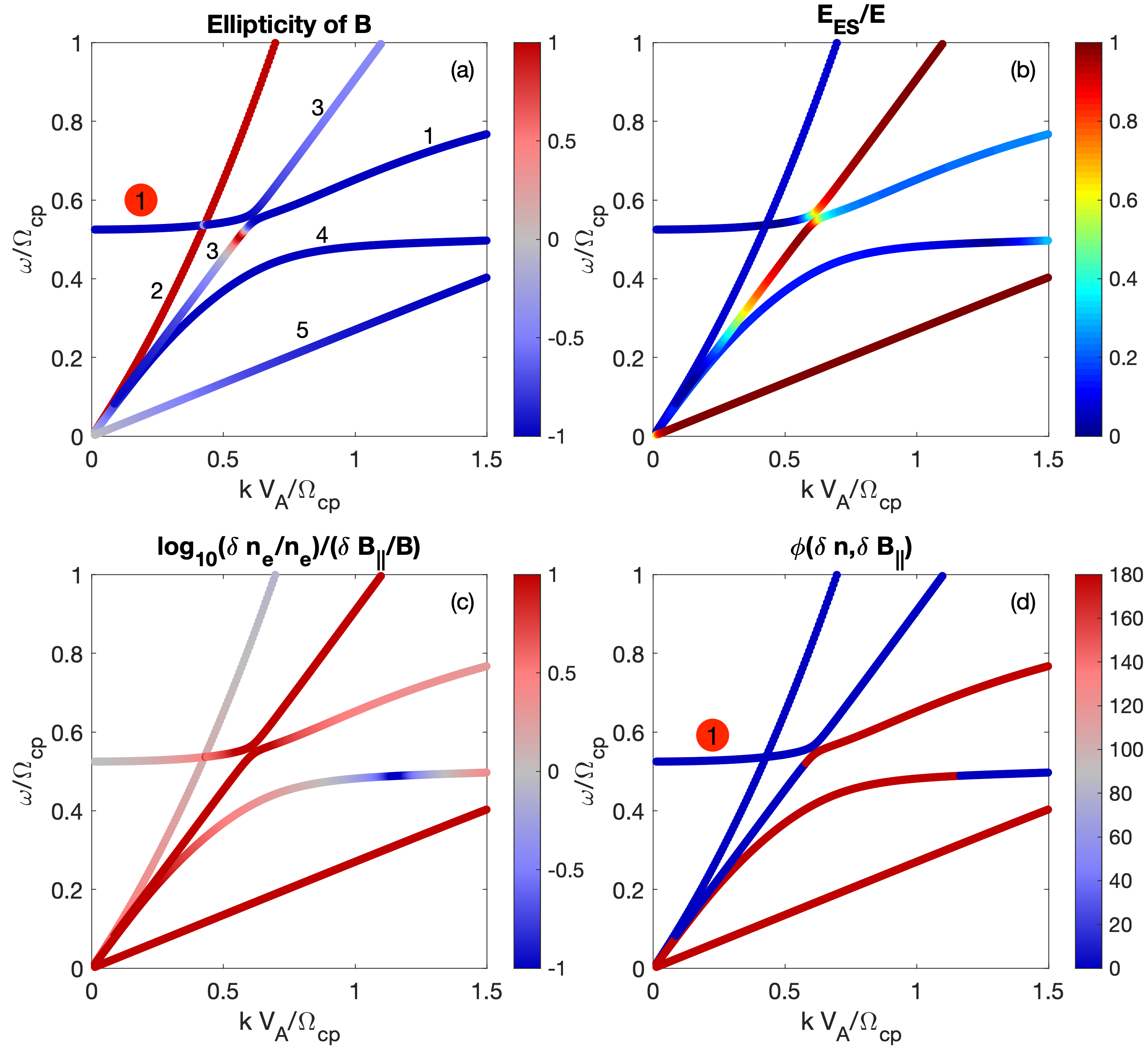}
   \caption{Dispersion relation for a 3-fluid plasma model for quasi-parallel propagation (wave-normal angle of $5^{\circ}$). Plasma parameters corresponding to Event 2 (discussed later) are $B_0 = 7$ nT, $n_e$ = 13 cm$^{-3}$, n$_p$ = 0.95 n$_e$, $n_{\alpha}$ = 0.025 n$_e$, $T_e = 10$ eV, $T_p = 5$ eV, $T_{\alpha} = 5$ eV.
              \label{fig:Wave_dispersion}}%
    \end{figure*}

Now that we have found that the observed waves exhibit frequencies close to $f_{cp}$, both right- and left-handed circular polarization, wave normal angles close to zero and $\xi_e \sim 1$, we use the thermal fluid plasma model to study the waves in this frequency range to identify the wave mode(s) corresponding to the observed waves. Our model consists of three fluids: electrons, protons, and alphas and we have assumed the polytropic indices are 1. The details of the model are given in Appendix A.

To illustrate a typical situation found in the solar wind, we show in Figure \ref{fig:Wave_dispersion} the dispersion relation in the plasma reference frame obtained using the 3-fluid model for plasma conditions of event 2 which we discuss in detail later. As we have no particle measurements available during the interval of interest we assume some typical values for the electron and ion temperatures which are given in the figure caption. We also assume that the alphas contain 10\% of the ion mass density. We consider a slightly oblique wave propagation with respect to the background magnetic field with the wave normal angle of 5$^{\circ}$. The quantities we plot (including $\xi_e$) have a weak dependence on the angle.

Compared to the three wave-mode branches for the case of a single ion population, with two ion populations, we obtain five wave-mode branches. Figure \ref{fig:Wave_dispersion}a shows the resulting dispersion branches colored by the wave ellipticity. The ellipticity of +1 corresponds to right-handed circular polarization, and of -1 to left-handed circular polarization, zero ellipticity corresponds to linear polarization. Only one of the branches (branch 2) is right-handed and this branch is a fast magnetosonic branch. The other four branches have predominantly left-handed polarization. Branch 5 is an electrostatic branch, and we will not consider it as the observed waves have a strong magnetic component. Branch 4 is the shear Alfv\'en wave at low $k$. Branch 3 is the slow/ion-acoustic branch. Finally, branch 1 is the proton-band electromagnetic ion cyclotron (PB-EMIC) wave. It is limited to the frequency range bounded by the cyclotron frequencies of alphas and protons, $\Omega_{c\mathrm{He}++} < \omega < \Omega_{c\mathrm{H}+}$. We will show that the low-$k$ part of branch 1 (marked by a red circle), i.e. the part for which $\delta n$ and $\delta B_{||}$ are in phase (panel d), is matching the observed wave properties for the two events we discuss in detail later.

As we observe waves with both senses of polarization, we need to look at additional quantities to narrow down the search. We can study the compressibility ratio $\xi_e$ and the phase between $\delta n$ and $\delta B_{||}$ shown in Figure \ref{fig:Wave_dispersion}c. The statistical distribution in Figure \ref{fig:Wave_stats} shows that the compressibility ratio is confined to a range of 0.5--1.5. In the log scale used in Figure \ref{fig:Wave_dispersion}c this corresponds to -0.3--+0.2, the grey part of the color scale. The phase observations suggest that $\delta n$ and $\delta B_{||}$ are close to being in phase for the observed waves. We have already discarded electrostatic branch 5. We can also disregard most of branch 3 (except for the low-wavenumber part), as it is both electrostatic and has a very high compressibility ratio. There are three branches that fully of partially satisfy our criteria: branch 2, low-wavenumber part of branch 1 approaching cut-off at $\Omega_{c\mathrm{He++}}$, and the high-wavenumber of branch 4. We note that such large wavenumbers will give a substantial Doppler shift, as we often find the wave propagating aligned or anti-aligned with the solar wind velocity when the magnetic field is predominantly radial. This can be also used as a diagnostic and we will use it later. 

\section{Wave identification}
   \begin{figure*}
   \centering
   \centering\includegraphics[scale=0.6]{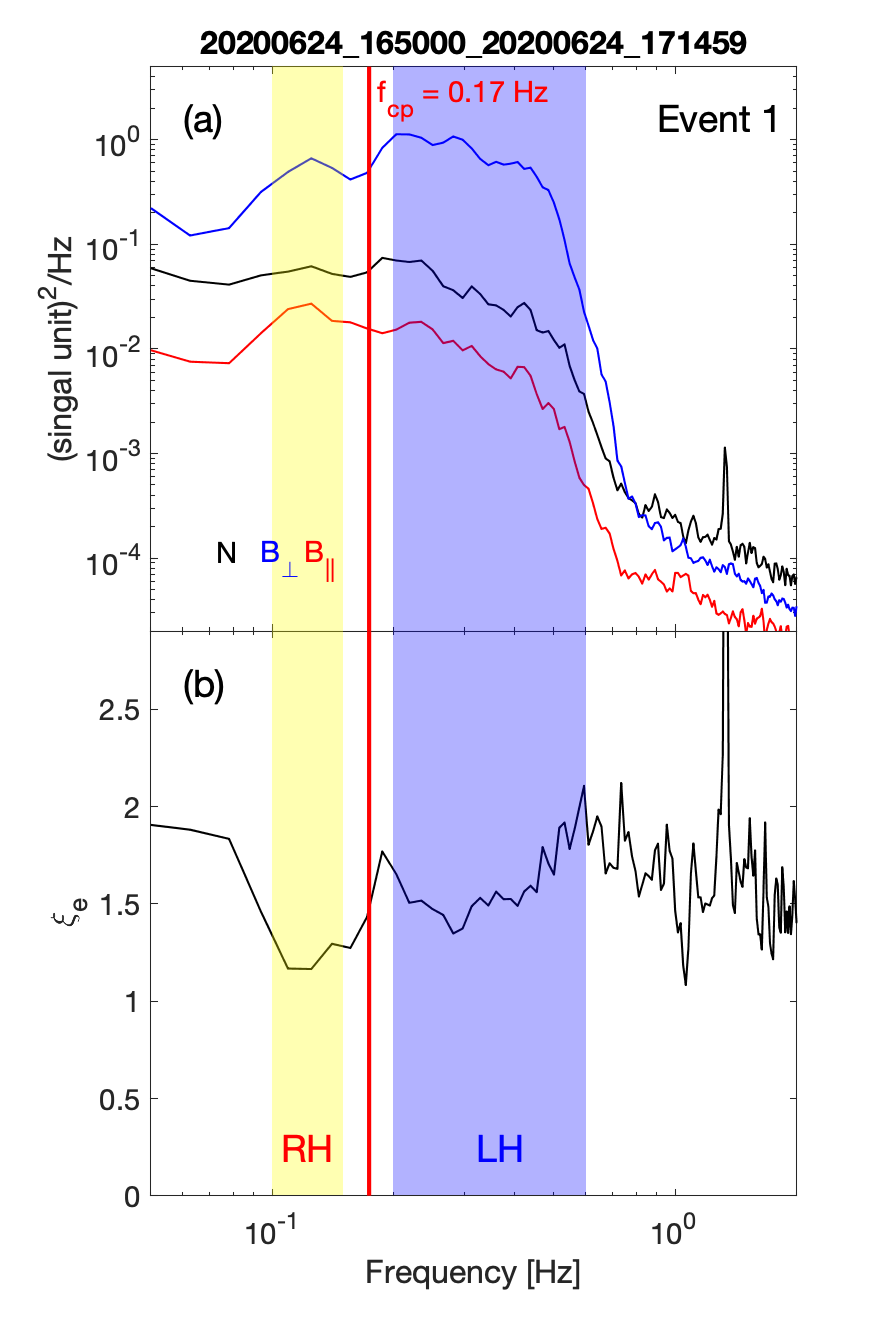}
   \centering\includegraphics[scale=0.6]{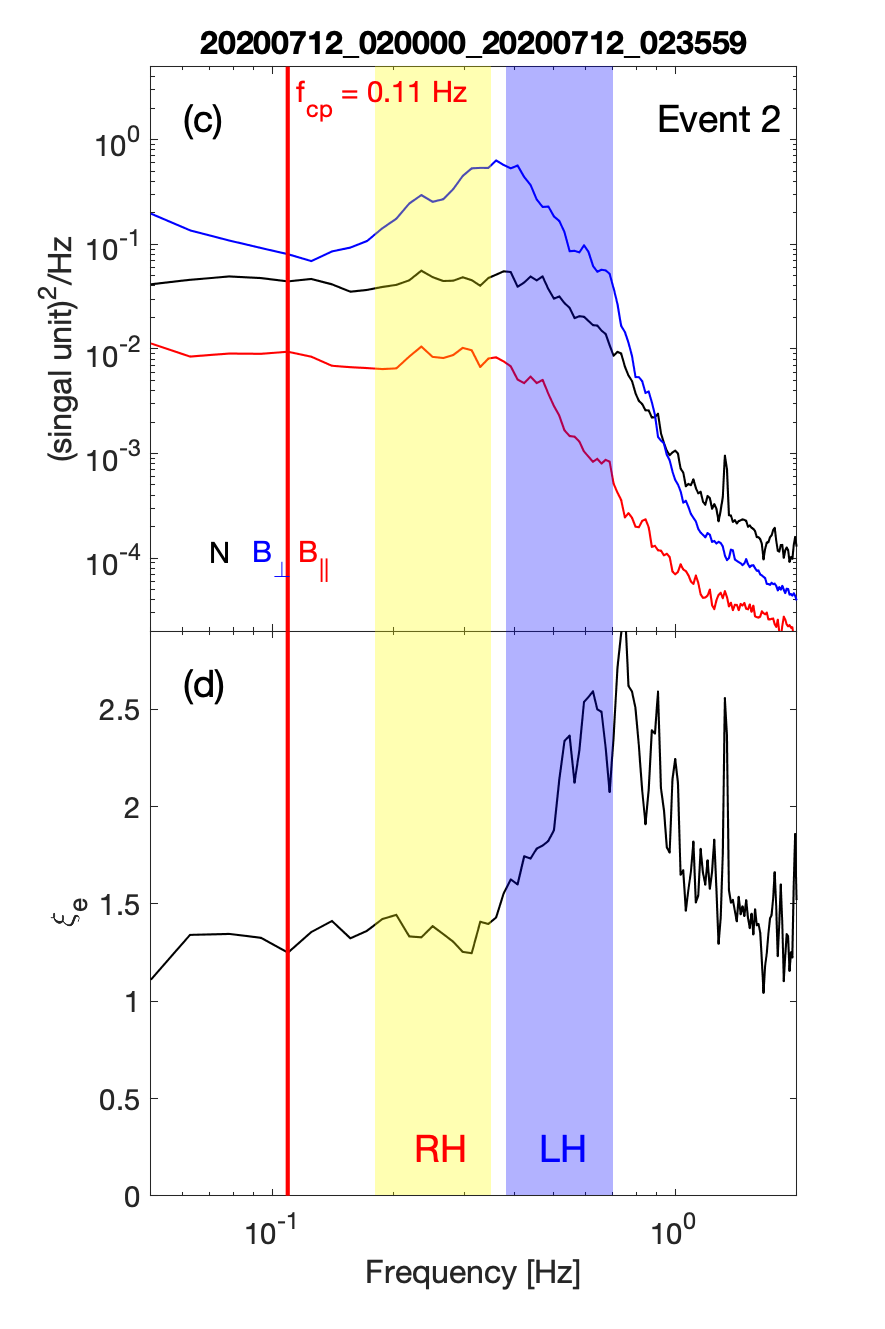}
   \caption{Spectra for two double-banded events. (a,c) Spectra of $n_e$, $B_{\perp}$ and $B_{||}$ (b,d) compressibility ratio $\xi_e)$ for the two events. The yellow and violet shadings show the frequency ranges for the right- (RH) and left-handed (LH) polarization respectively. The red vertical line shows the local proton cyclotron frequency. We note that a sharp peak in the  $n_e$ spectrum at $\sim$1.2 Hz is related to the spacecraft interference.
              \label{fig:Double_banded_examples}}%
    \end{figure*}

   \begin{figure*}
   \centering
   \includegraphics[width=\hsize]{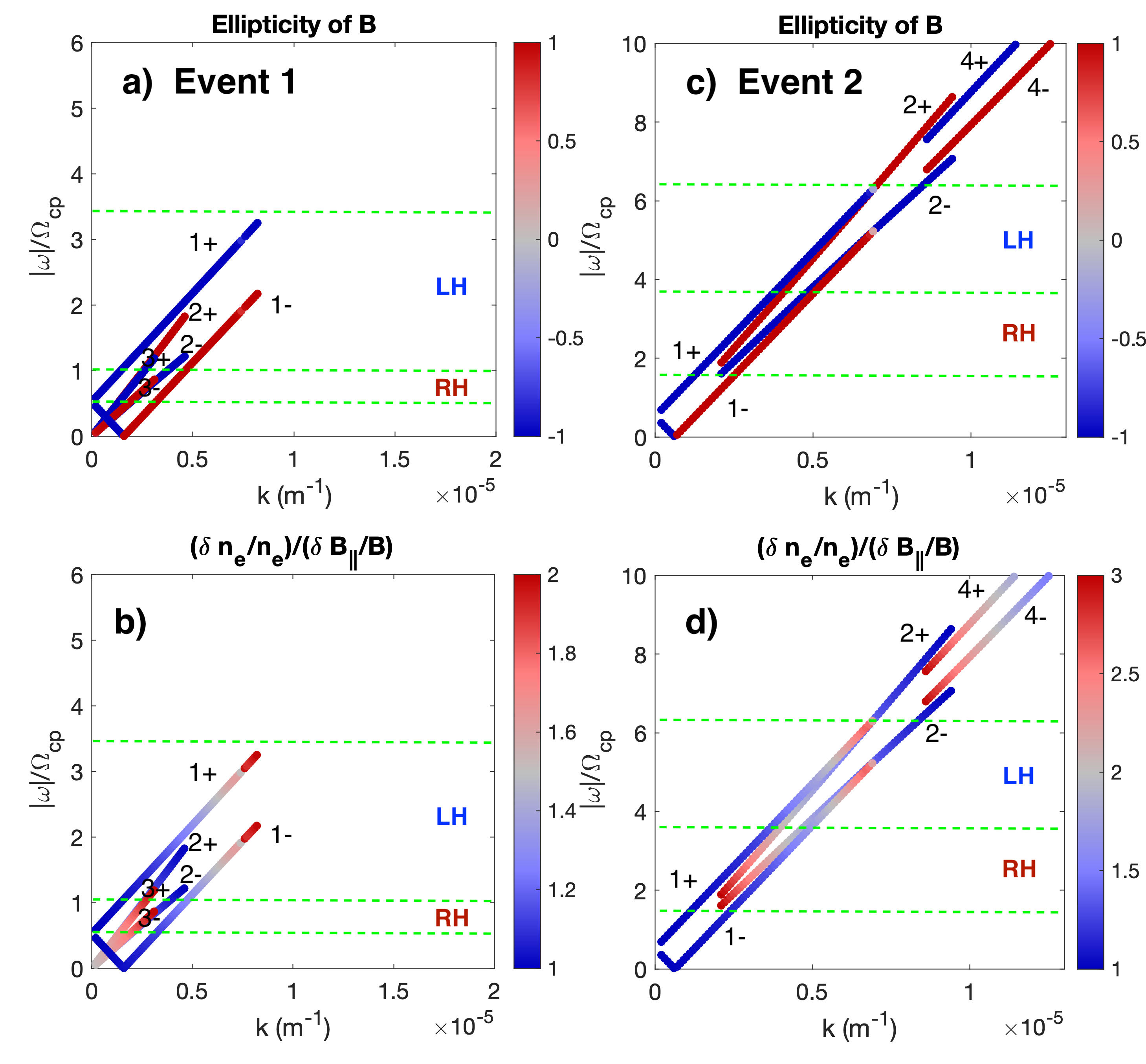}
   \caption{Doppler shifted dispersion relations. (a,c) show the sense of polarization and (b,d) compressibility ratio $\xi_e$. We mark the outward propagating waves by "+" and inward-propagating by "-". We also indicate the frequency ranges for the observed RH and LH bands (green dashed lines), so that we will be looking for the "blue" branches in the LH band, and the "red" branches in the RH band. 
              \label{fig:Doppler_shifted_spectra}}%
    \end{figure*}

Now we will consider two specific wave events and will determine the likely wave branches corresponding to the observations based on the comparison of the observed properties to the 3-fluid theory. For both events, we observed both right- and left-handed waves at different frequencies. Event (1) is on June 24, 16:50--17:15 UT. Figure \ref{fig:Event_overview}i shows that the waves are left-handed above $f_{cp}$ and right-handed below. We can also see that $\delta n$ and $\delta B_{||}$ are in phase at this time (Fig. \ref{fig:Event_overview}g). The event is observed within a slow solar wind interval, $V_{sw} \sim 380$ km/s. With an absence of ion data, we use de Hoffmann-Teller analysis to determine $V_{sw}$ \citep{steinvall_solar_2021}. The second event (event 2, observed on July 12, 02:00--02:36) is similar to the first one, but it is observed during the fast solar wind $V_{sw} \sim 560$ km/s and correspondingly we observe the waves of both polarizations Doppler-shifted to frequencies above $f_{cp}$.

Figure \ref{fig:Double_banded_examples} shows the power spectra of $N_{\mathrm{RPW}}$,  $B_{||}$ and $B_{\perp}$ in panels (a) and (c) and the compressibility ratio $\xi_e$ in panels (b) and (d) for the two events respectively. We indicate the frequency ranges for which the right-handed (RH, yellow) and left-handed (LH, violet) polarizations are observed. For event (1) the LH and RH bands are separated by $f_{cp}$, while for event (2) both bands are above $f_{cp}$ consistent with expected larger Doppler shift. We note that $\xi_e$ is increasing with frequency, and is contained to $2 > \xi_e > 1$ for Event (1) and $3 > \xi_e > 1$ for Event (2).

We use the observed ranges of $\xi_e$ to find the parts of the wave branches  which satisfy the observed ranges. We use the plasma parameters for each of the events to compute the wave properties. Where temperatures cannot be measured we have used nominal solar wind conditions. That should not have any major impact on the results as long as the observed interval is not very unusual in terms of the plasma temperature. We also compute the wave frequency in the spacecraft frame, i.e. we consider wave propagating both towards and away from the Sun and include the effect of the Doppler shift
\begin{equation}
   \omega = \omega_0 + \mathbf{k} \cdot \mathbf{V}_{\mathrm{sw}} \,,
\end{equation}
where $\omega_0$ is the wave frequency in the plasma frame, and $\mathbf{k}$ is the wave vector. For the outward-propagating waves, $\mathbf{k}$ has approximately the same direction as $\mathbf{V}_{\mathrm{sw}}$ and the dot products is positive leading to an increase of the spacecraft-frame frequency $\omega$. 

Figure \ref{fig:Doppler_shifted_spectra}a shows the dispersion relation for event (1) in the spacecraft frame with the color indicating the predicted handiness of polarization and the green dashed lines marking the frequency intervals where observed waves show right-handed (RH) and left-handed (LH) polarization. This allows one to identify the dispersion branches that satisfy the observed wave polarization. We plot only the parts of the dispersion branches satisfying the observed range of $\xi_e$ ($2 > \xi_e > 1$) that is satisfied by three dispersion branches. We can directly see that this provides a limitation of the range of the observed wavenumbers, $k < 8 \times 10^{-6}$~m$^{-1}$ corresponding to wavelength $\lambda > 10^6$~m. The numbers indicate the different branches and the outward- and inward-propagating waves are marked by "+" and "-" respectively.

We see that branch 1+ is the only one covering the entire range of the observed LH band $3.5 > \omega/\Omega_{cp} > 1$. This branch corresponds to branch 1 in Figure \ref{fig:Wave_dispersion}, specifically the low-$k$ part of it. There, the plasma frame frequency is approximately constant and the spacecraft-frame dispersion is close to linear, attributed primarily to the Doppler shift. The expected ranges of $\xi_e$ are shown in Figure \ref{fig:Doppler_shifted_spectra}b. We can see that the model $\xi_e$ for branch 1+ is in good agreement with the observations (Figure \ref{fig:Double_banded_examples}b), both in terms of the observed values and that $\xi_e$ is increasing with frequency. We conclude that branch 1+ is the only one satisfying the properties of the observed LH band.

Now we look at the RH frequency band. The inward-propagating waves of branch 1 (marked 1-) stay left-handed only for the smallest $k$ values and then switch their polarization to right-handed when the Doppler shift becomes large and the spacecraft-frame frequency $\omega$ becomes negative. The other right-handed branches in the RH frequency band are 2+ and 3-.   Predicted $\xi_e \sim 1.8$ for branch 3- is too high with respect to the observed $\xi_e \sim 1.2$. For branches 1- and 2+ the predicted $\xi_e$ is in good agreement with the observed $\xi_e$. So both branches 1- (inward-propagating PB-EMIC wave) and 2+ (outward-propagating fast magnetosonic wave) match the observed wave properties in the RH band.

Now we perform a similar analysis for event (2) for which the dispersion relations colored by the handedness of polarization and $\xi_e$ are shows in Figure \ref{fig:Doppler_shifted_spectra}c and d. As the solar wind speed is higher for this event, the waves are Doppler shifted to higher frequencies compared to event (1). We compare the predicted wave branches to the observed frequency range of the LH and RH bands. We see that branches 4+ and 4- (corresponding to branch 4 in Fig. \ref{fig:Wave_dispersion}) are outside of the observed frequency range due to the large $k$ and thus large Doppler shift. In the RH band, both branches 1- and 2+ are right-handed and match the observed frequencies. But if we take into account the observed compressibility ratio $\xi \sim 1.3$ (Fig. \ref{fig:Double_banded_examples}d), we see from Fig. \ref{fig:Doppler_shifted_spectra}d that for branch 2+ $\xi$ is higher than the observed value, and only branch 1- agrees well with the observations. Similarly, in the LH band, we are limited to branches 1+ and 2- based on polarization. However, looking at the observed $\xi$, which increases with frequency from 1.5 to 2.5, we find that only branch 1+ is in good agreement with the observations, and in particular, it features a clear increase of $\xi$ with frequency. Thus we find that for both LH and RH bands the observed waves correspond to the outward and inward propagating PB-EMIC wave (branches 1+ and 1-), while the fast magnetosonic waves (branches 2- and 2+) can be ruled out.   

We have considered two events for which we simultaneously observe RH and LR waves in adjacent frequency bands. For event (2) we find that both bands can be uniquely identified as the PB-EMIC wave with relatively small wavenumbers $k < 7 \times 10^{-6}$~m$^{-1}$. For such small wavenumbers, $k V_A/\Omega_{cp} < 1$, a drift of alpha particles with respect to protons does not lead to a significant modification of the dispersion relation and wave properties \citep{gomberoff_resonant_1991}. For such wavenumbers the PB-EMIC approaches the cut-off at $f_{c \mathrm{He++}}$, so the group velocity of these waves is low, so it is likely that we are observing them close to the source region. This is also consistent with both bands being observed simultaneously; one of the likely sources of the waves is the ion temperature anisotropy \citep{davidson_electromagnetic_1975}, which will generate waves propagating in both the inward and outward directions with similar wavenumber ranges. 
The inward- and outward-propagating waves, if generated in the same region, would propagate out of the generation region in different directions. So, after some time the waves will propagate apart, and we would be observing only the inward- or the outward-propagating waves in a given region. And the fact that we observe both directions simultaneously indicates that we are observing the source region.

As one can see from Figure \ref{fig:Doppler_shifted_spectra} there will be an overlap between the outward- (LF) and inward-propagating (RH) waves, in a certain frequency (spacecraft frame) range. This is consistent with the observed gap in the coherence between the two bands (Figure \ref{fig:Event_overview}h), but no gap in the power spectrum. For event (1) we also uniquely identify the PB-EMIC wave as the LH mode, while for the RH wave the identification does not yield a unique branch. However, as the observed RH and LH waves have very similar properties, i.e. amplitude, frequency, wave-normal angle, it is natural to assume that even for that event the same wave mode (PB-EMIC) is providing both the LH and RH observed waves. It is more likely that these two belong to the same wave mode than to the different modes, as the growth of different modes will normally have different growth rates resulting in a difference in the resulting wave properties. Overall, we conclude that such double-banded events likely correspond to the wave source region, where a particular instability, such as the ion temperature anisotropy instability, is generating waves in both directions.

\section{Conclusions}

We present observations of plasma turbulence and quasi-circularly-polarized electromagnetic waves close to proton-cyclotron frequency $f_{cp}$ near the first perihelion encounter by Solar Orbiter ($\sim$0.5~AU from the Sun) which we analyze using the magnetic field and plasma density measurements. The key results are as follows:

\begin{enumerate}
   \item We present the density calibration based on the probe-to-spacecraft potential and QTN measurements by RPW. We then use the obtained density $N_{\mathrm{RPW}}$ to compare the power spectra and time-series for density fluctuations to fluctuations of the magnetic field in two samples of solar wind. The observed spectral slopes of the transverse magnetic component are compatible with standard turbulence, and are the same for the studied turbulent intervals. 
   On the other hand, for the parallel magnetic field component and density fluctuations the spectral slope is fully compatible with turbulence in interval (I), where almost no waves are observed, but it is shallower in interval (II), rich in waves, suggesting the poorly developed nature of the turbulence.
   We also find that the density fluctuations are out-of-phase with the compressible component of magnetic fluctuations for intervals of turbulence, which is consistent with earlier analysis based on particle data at lower frequencies (larger spatial scales). These analyses show that RPW provides high-quality and high-cadence measurements of plasma (electron) density. The absolute accuracy of this measurement is assessed using the observations of Langmuir waves by \citet{graham_kinetic_2021}.
   \item We further investigate the quasi-circular electromagnetic waves close to the proton cyclotron frequency, $f_{cp}$. These waves have wave-normal angles (with respect to background B) close to zero and have either left- or right-handed polarization in the spacecraft frame. Despite the small wave-normal angles, these waves have a detectable compressible component, which we find to be in phase with fluctuations in density. We investigate statistically the compressibility ratio $\xi_e = (\delta n/n)/(\delta B_{||}/B)$ for the observed waves and find a rather narrow distribution confined to a range $0.5 > \xi_e > 1.5$.
   \item We analyze in detail two specific events with the simultaneous presence of left- or right-handed waves at different frequencies. We compare observed wave properties such as the frequency ranges for the two senses of polarization and compressibility ratio $\xi_e$, and in particular the dependence of $\xi_e$ on frequency, to a prediction of the three-fluid (electrons, protons, and alphas) model. We take into account the Doppler shift changing the wave frequency observed in the spacecraft frame depending on the solar wind speed and the wave properties. From this we can find a limit on the observed wavenumbers, $10^{-6} < k < 7 \times 10^{-6}$~m$^{-1}$, which corresponds to wavelength $7 \times 10^6 >\lambda > 10^6$~m. We conclude that it is most likely that both the left- and right-handed waves correspond to proton-band electromagnetic ion cyclotron (PB-EMIC) waves propagating in the outwards and inwards directions respectively. 
   \item The fact that both wave polarizations are observed at the same time and that the PB-EMIC waves have a low group velocity for the observed range of $k$ (close to the cut-off at the $\alpha$-cyclotron frequency) suggests that the double-banded events occur in the source regions of the waves. A likely source of such waves is an ion temperature anisotropy instability generating waves in both the field-aligned and the opposite directions, which makes this wave important for understanding solar wind heating. The quasi-circular waves near $f_{cp}$ are common for the intervals with radial magnetic field \citep{jian_observations_2010, bale_highly_2019} which likely contain fresh solar wind emerging from coronal holes \citep{smith_ulysses_1995}, and it is likely that such waves are one of the primary mechanisms controlling ion dynamics in absence of significant levels of turbulence in such regions. Further studies including analysis of the ion distributions observed by SWA-PAS are necessary to identify the details of the instability. 
\end{enumerate}

  The presented first results show that the plasma density based on RPW measurements provides an excellent opportunity to study density fluctuations in the solar wind at fast temporal scales, and which combined with other observations by Solar Orbiter will help us to improve our understanding of the solar wind physics.

\begin{acknowledgements}
Solar Orbiter is a space mission of international collaboration between ESA and NASA, operated by ESA.
  	We thank the entire Solar Orbiter team and instrument PIs for data access and support. Solar Orbiter data are available at http://soar.esac.esa.int/soar/\#home. This work is supported by the Swedish Research Council, grant 2016-05507, and Swedish National Space Agency (SNSA) grants 20/136 and 128/17. LSV was funded by the Swedish Contingency Agency grant 2016-2102 and by SNSA grant 86/20. CNES and CDPP are acknowledged for the support to the French co-authors. Solar Orbiter magnetometer operations are funded by the UK Space Agency (grant ST/T001062/1). Tim Horbury is supported by STFC grant ST/S000364/1.

\end{acknowledgements}

%
%
\bibliographystyle{aa}

\appendix \label{app1}
\section{Fluid model of plasma waves} \label{app1}
To derive the dispersion equation for thermal plasma waves we start with the usual fluid and Maxwell's equations:
\begin{equation}
\frac{\partial n_j}{\partial t} + \nabla \cdot (n_j {\bf V}_j) = 0,
\label{conteq}
\end{equation}
\begin{equation}
m_j n_j \left[ \frac{\partial {\bf V}_j}{\partial t} + ({\bf V}_j \cdot \nabla) {\bf V}_j \right] = \epsilon_j e n_j ({\bf E} + {\bf V}_j \times {\bf B}) - \nabla P_j,
\label{mom}
\end{equation}
\begin{equation}
P_j = P_{0,j} \left( \frac{n_j}{n_{0,j}} \right)^{\gamma_j}
\label{closure}
\end{equation}
\begin{equation}
\nabla \cdot {\bf E} = \frac{\rho}{\epsilon_0},
\label{max1}
\end{equation}
\begin{equation}
\nabla \cdot {\bf B} = 0,
\label{max2}
\end{equation}
\begin{equation}
\nabla \times {\bf E} + \frac{\partial {\bf B}}{\partial t} = 0
\label{max3}
\end{equation}
\begin{equation}
\nabla \times {\bf B} - \frac{1}{c^2} \frac{\partial {\bf E}}{\partial t} = \mu_0 {\bf J},
\label{max4}
\end{equation}
where $n$ is the number density, ${\bf V}$ is the bulk velocity, $P$ is the scalar pressure, $\gamma$ is the polytropic index, $m$ is the particle mass, ${\bf E}$ is the electric field, ${\bf B}$ is the magnetic field, $\rho$ is the charge density, ${\bf J}$ is the current density, $e$ is the magnitude of the unit charge, $c$ is the speed of light, and $\epsilon_0$ and $\mu_0$ are the permittivity and permeability of free space. The subscripts $j$ refer to the particle species, and $\epsilon_j$ is $+1$ for protons, $+2$ for alphas, and $-1$ for electrons. 

We divide the fields and particle moments into fluctuation and non-fluctuating quantities 
${\bf Q} = {\bf Q}_0 + \delta {\bf Q}$, and without loss of generality assume ${\bf B}_0 = (0, 0, B_0)$ and 
the wave vector is given by ${\bf k} = (k_x, 0, k_z)$. We assume $V_{0,j} = 0$, although we note that the proton and alpha bulk velocities can differ in the solar wind \citep{marsch_solar_1982}. This can modify the dispersion relations for large $k$. We assume a plane wave solution 
of the form
\begin{equation}
\delta {\bf Q}(\omega,{\bf k}) = \delta {\bf Q} \exp{(-i \omega t + i k_x x + i k_z z)},
\label{quant}
\end{equation}
where $\omega$ is the angular frequency, and substitute into equations (\ref{conteq})--(\ref{max4}). 
From Maxwell's equations we obtain the usual wave equation 
\begin{equation}
{\bf n} \times {\bf n} \times \delta {\bf E} + {\bf K} \cdot \delta {\bf E} = 0, 
\label{waveeq}
\end{equation}
where ${\bf n} = c {\bf k}/\omega$ is the refractive index, ${\bf K} = {\bf I} + i {\bf \sigma}/\epsilon_0\omega$ is the dimensionless dielectric tensor, and ${\sigma}$ is the conductivity tensor, which is given by $\delta {\bf J} = {\sigma} \cdot \delta {\bf E}$. 

From equation (\ref{conteq}) we obtain
\begin{equation}
\delta n_j = \frac{n_j}{\omega} \left( k_x \delta V_{x,j} + k_z \delta V_{z,j} \right).
\label{conteq2}
\end{equation}

The fluctuating velocities are obtained by linearizing equation (\ref{mom}) and using equations (\ref{closure}) and 
(\ref{conteq}) to eliminate $\delta P_j$ and $\delta n_j$, respectively. After some tedious but straightforward calculations we obtain:
\begin{strip}
\begin{equation}
\delta V_{x,j} = \frac{e}{m_j D_j} \left( i \epsilon_j \omega (\omega^2 - \gamma_j v_j^2 k_z^2) \delta E_x - 
\Omega_{cj} (\omega^2 - \gamma_j v_j^2 k_z^2) \delta E_y + i \epsilon_j \omega \gamma_j v_j^2 k_x k_z \delta E_z \right),
\label{Vx}
\end{equation}
\begin{equation}
\delta V_{y,j} = \frac{e}{m_j D_j} \left( \Omega_{cj} (\omega^2 - \gamma_j v_j^2 k_z^2) \delta E_x + 
i \epsilon_j \omega (\omega^2 - \gamma_j v_j^2 k^2) \delta E_y + \Omega_{cj} \gamma_j v_j^2 k_x k_z \delta E_z \right),
\label{Vy}
\end{equation}
\begin{equation}
\delta V_{z,j} = \frac{e}{m_j D_j} \left( i \epsilon_j \omega \gamma_j v_j^2 k_x k_z \delta E_x - 
\Omega_{cj} \gamma_j v_j^2 k_x k_z \delta E_y + i \epsilon_j \omega (\omega^2 - \Omega_{cj}^2 - \gamma_j v_j^2 k_x^2) \delta E_z \right),
\label{Vz}
\end{equation}
\end{strip}
where 
\begin{equation}
D_j = \omega^2 (\omega^2 - \gamma_j v_j^2 k^2) - \Omega_{cj}^2 (\omega^2 - \gamma_j v_j^2 k_z^2), 
\label{Dj}
\end{equation}
and $\Omega_{cj}$ is the angular cyclotron frequency of species $j$, $v_j = \sqrt{k_B T_j/m_j}$ is the thermal 
speed, $T_j$ is the scalar temperature, and $k_B$ is Boltzmann's constant.
The fluctuating current density is given by $\delta {\bf J} = \sum_j \epsilon_j e n_j \delta {\bf V}_j$, whence we 
obtain ${\bf \sigma}$ and ${\bf K}$ using equations (\ref{Vx})--(\ref{Dj}).

The dispersion equation is given by: 
\begin{equation}
\begin{pmatrix}
K_{xx} - n_z^2 & K_{xy} & K_{xz} + n_x n_z \\
- K_{xy} & K_{yy} - n^2 & K_{yz} \\
K_{xz} + n_x n_z & - K_{yz} & K_{zz} - n_x^2 
\end{pmatrix}
\begin{pmatrix}
\delta E_x \\
\delta E_y  \\
\delta E_z 
\end{pmatrix}
= 0,
\label{waveeq}
\end{equation}
where 
\begin{equation}
K_{xx} = 1 - \sum_j \frac{\omega_{pj}^2 (\omega^2 - \gamma_j v_j^2 k_z^2)}{D_j},
\label{Kxx}
\end{equation}
\begin{equation}
K_{xy} = -\sum_j i \epsilon_j \frac{\omega_{pj}^2 \Omega_{cj} (\omega^2 - \gamma_j v_j^2 k_z^2)}{\omega D_j},
\label{Kxy}
\end{equation}
\begin{equation}
K_{xz} = -\sum_j \frac{\omega_{pi}^2 \gamma_j v_j^2 k_x k_z}{D_j},  
\label{Kxz}
\end{equation}
\begin{equation}
K_{yy} = 1 - \sum_j \frac{\omega_{pj}^2 (\omega^2 - \gamma_j v_j^2 k^2)}{D_j},
\label{Kyy}
\end{equation}
\begin{equation}
K_{yz} = \sum_j i \epsilon_j \frac{\Omega_{cj} \omega_{pj}^2 \gamma_j v_j^2 k_x k_z}{\omega D_j},
\label{Kyz}
\end{equation}
\begin{equation}
K_{zz} = 1 - \sum_j \frac{\omega_{pj}^2 (\omega^2 - \Omega_{cj}^2 - \gamma_j v_j^2 k_x^2)}{D_j},
\label{Kzz}
\end{equation}
Here $\omega_{pj}$ is the plasma frequency of species $j$. 

The wave equation can be written as: 
\begin{multline}
(K_{xx}^2 - n_z^2) [(K_{yy} - n^2)(K_{zz} - n_x^2) + K_{yz}^2] 
+ K_{xy} [ K_{xy} (K_{zz} - n_x^2) + K_{yz} (K_{xz} + n_x n_z)] \\
+ (K_{xz} + n_x n_z)[ K_{xy} K_{yz} - (K_{yy} - n^2)(K_{xz} + n_x n_z)] = 0.
\label{dispeq}
\end{multline}

Once the solutions are found the relative magnitudes and phases of the components of $\delta {\bf E}$ are 
given by equation (\ref{waveeq}). The magnetic field fluctuations $\delta {\bf B}$ are given by Faraday's law [equation (\ref{max3})], $\delta {\bf V}_j$ are given by equations equations (\ref{Vx})--(\ref{Dj}),  and 
$\delta n_j$ are given by equation (\ref{conteq2}).

\end{document}